\pdfoutput=1

\IfFileExists{latexml.sty}{\RequirePackage{latexml}}{\newif\iflatexml\latexmlfalse}

\iflatexml %only 
\else %not latexml
\fi %return to latex all

\documentclass[conference]{IEEEtran}  

\usepackage[utf8]{inputenc}
\usepackage[T1]{fontenc}

\def\equalizetitlepagecols {0mm} % OR \global\def\equalizetitlepagecols {-8mm}
\IEEEoverridecommandlockouts

\usepackage{amsmath,amsfonts} \usepackage{array}
\iflatexml %only
\else %not latexml
\usepackage{tgcursor}
\usepackage{algorithm} % Banned IEEE formats figure 1 code better w/ algorithm
\fi %return to latex all

\usepackage{comment}
\usepackage{alltt} 
\usepackage{upquote} % Provide textasciigrave etc despite T1 encoding

\usepackage{graphicx} % rotated boxes in table

\usepackage{amsthm} % Theorem statements in italics

\newlength{\bufferhline} 

\def\myblue{blue}
\def\myblack{black}

\iflatexml %only
\usepackage{subcaption}
\usepackage{color}
\usepackage{soul}
\setulcolor{blue}
\usepackage[colorlinks=true]{hyperref}
\usepackage[square,numbers]{natbib}
\newcommand{\shortdoi}[1]{\href{https://doi.org/#1}{\ul{\color{\myblack}\texttt{doi:#1}}}}
\newcommand{\wvcite}[1]{\citep[{\color{\myblue}{$\Rightarrow$}}][]{#1}}
\newcommand{\wvref}[1]{\hyperref[#1]{{\color{\myblue}{$\Rightarrow$}}{\color{\myblack}{\ref{#1}}}}}
\newcommand{\wveqref}[1]{\hyperref[#1]{{\color{\myblue}{$\Rightarrow$}}{\color{\myblack}{\eqref{#1}}}}}
\def\mybegineqstar#1{\begin{equation}\label{#1}}
\def\myendeqstar{\end{equation}}
\def\mytag#1{}
\else %not latexml
\usepackage{hhline}
\usepackage[normalem]{ulem}
\usepackage[svgnames]{xcolor}
\usepackage[colorlinks=true,allcolors=\myblack]{hyperref}
\usepackage[table]{hypcap}
\newcommand{\shortdoi}[1]{{\color{\myblue}\uline{\color{\myblack}\href{https://doi.org/#1}{\texttt{doi:#1}}}}}
\newcommand{\wvcite}[1]{{\color{\myblue}\uline{{\color{\myblack}\cite{#1}}}}}
\newcommand{\wvref}[1]{{\color{\myblue}\uuline{{\color{\myblack}\ref{#1}}}}}
\newcommand{\wveqref}[1]{{\color{\myblue}\uline{{\color{\myblack}\eqref{#1}}}}}
\def\mybegineqstar#1{\hbox{\refstepcounter{equation}\label{#1}}\begin{equation*} }
\def\myendeqstar{\end{equation*}}
\def\mytag#1{\tag{#1}}
\fi %return to latex all

\newcommand \doichoice [1] {\shortdoi{#1}} % Short Doi with hyperlinks required editing in this .tex

\iflatexml %only
\else %not latexml
\usepackage{afterpage}
\usepackage{footnotehyper}
\fi %return to latex all

\newcommand \nbdd {\nobreakdash}

\newlength{\myrowskip}

\iflatexml %only
\newcommand{\myTwoColCaptionvspace} [1][]{}
\newcommand{\myTwoColSectionvspace} [1][]{} 
\newcommand{\myTwoColTableLiftvspace} [1][]{}
\newcommand{\mydepthbox}{\vrule height0pt depth10pt width0pt}

\def\mytabcolsepA{2.0em}
\else %not latexml
\newcommand{\myTwoColCaptionvspace} [1][-2.0mm]{\vspace{#1}}
\newcommand{\myTwoColSectionvspace} [1][-2.0mm]{\vspace{#1}} % Needed with amsthm 2-column 
\newcommand{\myTwoColTableLiftvspace} [1][-3.0mm]{\vspace{#1}} % Needed with amsthm 2-column
\newcommand{\mydepthbox}{\vrule height0pt depth0pt width0pt}

\def\mytabcolsepA{0.15em}
\fi %return to latex all

\iflatexml %only

\newtheorem{theorem}{Theorem}
\newtheoremstyle{mystyle}% 
  {}%                                     % Space above
  {}%                                     % Space below
  {\noindent\itshape}%           % Body font
  {}%                                     % Indent amount
  {\bfseries\large}%               % Theorem head font
  {~:}%                                   % Punctuation after theorem head
  {\newline}%                         % Space after theorem head, ' ', or \newline
  {\noindent\thmname{#1}\thmnumber{ #2}\thmnote{ (#3)}} % Theorem head spec (empty = normal)
\theoremstyle{mystyle}
\newtheorem{theorem}{Theorem}
\newtheorem{algorithm}{Algorithm(s)}

\def\myabstract{myabstractenv}
\def\mykeywords{mykeywordsenv}

\else %not latexml

\def\myabstract{abstract}
\def\mykeywords{IEEEkeywords}

\fi %return to latex all

\def  \BeginAlgFigure {\begin{algorithm}}
\def  \EndAlgFigure {\end{algorithm}}

\usepackage[english]{babel} % \usepackage{doi} redundant

\makeatletter
\let\ORIbbl@fixname\bbl@fixname
\def\bbl@fixname#1{%
  \@ifundefined{languagealias@\expandafter\string#1}
    {\ORIbbl@fixname#1}
    {\edef\languagename{\@nameuse{languagealias@#1}}}%
}
\newcommand{\definelanguagealias}[2]{%
  \@namedef{languagealias@#1}{#2}%
}
\let\BIBforeignlanguage\foreignlanguage
\makeatother

\definelanguagealias{en}{english}

\begin{document}
% \linepenalty=290 % changes hyphenation fo possibly shorten paper
\iflatexml %only
\def\mythankstext{\copyright The author, dunning@bgsu.edu, %
grants copyright permissionsas specified in\\
Creative Commons License CC BY-SA 4.0. (See Endnote 2)\\ 
\\
}
\else %not latexml
\def\mythankstext{\copyright The author  grants copyright permissions as specified in %
Creative Commons License CC BY-SA 4.0.  Attributions should reference this manuscript. %
Version of \today.
}
\fi %return to latex all
\newcommand \thankstext {\mythankstext}

\title{\LARGE Totally Disjoint 3-Digit Decimal Check Digit Codes \\[-1.2ex]}

\author{\IEEEauthorblockN{Larry A. Dunning%
\iflatexml %only 
\\[0.5ex]%
\else %not latexml, 
, Prof. Emeritus}%
\fi %return to latex all
\thanks{\thankstext}
\IEEEauthorblockA{\small Dept. of Computer Science\\
Bowling Green State University\\
Bowling Green, Ohio 43403, USA\\
Email: dunning@bgsu.edu} \\[-3.0ex]} 

\makeatletter
\def\myDoublePARstart{\@IEEESAVECMDIEEEPARstart}
\makeatother

\maketitle

\iflatexml %only
\else %not latexml
% Add page numbers to author's taste
\makeatletter
\let\lastpagenumber2 % Next line works for newer versions
\def\ps@plain{%
  \let\@mkboth\@gobbletwo
   \renewcommand{\@oddhead}{}%
  \renewcommand{\@evenhead}{}%
  \renewcommand{\@evenfoot}{\reset@font\small\hfil\thepage{}~of~\lastpagenumber \hfil}%
  \renewcommand{\@oddfoot}{\@evenfoot}%
}
\pagestyle{plain} % apply the changes
\makeatother
\thispagestyle{plain}
\fi %return to latex all

\begin{\myabstract}
In 1969 J. Verhoeff provided the first examples of a decimal error detecting code 
using a single check digit to provide protection  against all single, 
transposition and adjacent twin errors. 
The three versions of such a code that he presented are length 3\nbdd-digit codes with 2 information digits. 
Existence of a 4\nbdd-digit code would imply the existence of 10 such disjoint 3\nbdd-digit codes. 
This paper presents 3 pairwise disjoint 3\nbdd-digit codes. 
The codes developed herein, have the property that the knowledge of the multiset of digits included in a word  is
sufficient to determine the entire codeword even though their positions were unknown. 
Thus the codes are permutation-free, and this fulfills his desire to eliminate cyclic errors. 
Phonetic errors, where 2 digit pairs of the forms X0 and 1X are interchanged, 
are also eliminated.  
\end{\myabstract}

\begin{\mykeywords}
Decimal error detection, disjoint coding, transposition errors, 
twin errors, phonetic errors, permutation-free.
\end{\mykeywords}

\enlargethispage{\equalizetitlepagecols}

\myTwoColSectionvspace[-6mm]
\section{Verhoeff's 3-digit Decimal Codes}
\iflatexml %only
\raggedleft{In
\else %not latexml 
\myDoublePARstart{I}{n} 
\fi %return to latex all
his 1969 monograph \mbox{Jacobus Verhoeff~\wvcite{Verhoeff1969}} presented %
three variations of a ``curious 3\nbdd-digit decimal code", derived from a block design. %
The arguably best of these is shown in \mbox{Table~\wvref{VerhoeffIrregular}.} %
Each table entry \mbox{$S(r,c)$} gives the middle digit, \texttt{s}, of the codeword \texttt{(rsc)} throughout this paper, allowing a simpler correspondence between properties of the code table %
and the requirements for detecting various error types. These are detailed in \mbox{Table~\wvref{ErrorTypes}.} %
All three codes contained all the triple codewords of the form \texttt{(xxx)}, %
as is reflected by the main diagonal of \mbox{Table~\wvref{VerhoeffIrregular},} and do not detect triple errors. %
The code shown in \mbox{Table~\wvref{VerhoeffIrregular}} is preferred because it catches all but 16 cyclic errors %
and detects phonetic errors. %
The 3 pairwise disjoint codes developed in \mbox{Section~\wvref{permfree}} have the %
\textit{permutation-free} property.
\myTwoColSectionvspace[-4.0mm]
\section{Disjoint Codinng}
On page 19 of his seminal manuscript  \mbox{J. Verhoeff~\wvcite{Verhoeff1969}} noted that multiple pairwise-disjoint 
check-digit codes could be of use when multiple organizations ``like different bank branches'' utilize the same 
system of assignment for e.g. account numbers.  
The identification of the branch then serves as an additional check while account numbers would be unique across the system.
For the three disjoint codes developed here, 100 lockers at each of three locations 
would have 300 uniquely numbered lockers.
\iflatexml %only
{
\centering%
\myTwoColCaptionvspace
\setlength{\tabcolsep}{0.5em}
\begin{table}[!h]\label{VerhoeffIrregular}
\myTwoColTableLiftvspace
\caption{Verhoeff's Irregular Code}
\centering
\myTwoColCaptionvspace
\setlength{\tabcolsep}{0.5em}
\begin{tabular}{c|cccccccccc}
$S$&0&1&2&3&4&5&6&7&8&9\\
\hline
\rule{0pt}{2ex}   
0&0&3&4&9&6&7&5&8&2&1\\
1&5&1&0&2&8&3&9&6&7&4\\
2&7&6&2&4&1&0&8&9&3&5\\
3&1&5&8&3&7&6&4&0&9&2\\
4&2&9&7&5&4&8&1&3&0&6\\
5&6&7&9&0&3&5&2&4&1&8\\
6&3&8&1&7&5&9&6&2&4&0\\
7&9&4&5&8&2&1&0&7&6&3\\
8&4&0&6&1&9&2&3&5&8&7\\
9&8&2&3&6&0&4&7&1&5&9\\
\end{tabular}
\myTwoColCaptionvspace[-4.0mm]
\end{table}
}}
\else %not latexml 
{
\centering%
\myTwoColCaptionvspace
\setlength{\tabcolsep}{0.5em}
\begin{table}[!h]\label{VerhoeffIrregular}
\myTwoColTableLiftvspace
\caption{Verhoeff's Irregular Code}
\centering
\myTwoColCaptionvspace
\setlength{\tabcolsep}{0.5em}
\framebox(150,110){\par
\begin{tabular}{c|cccccccccc}
$S$&0&1&2&3&4&5&6&7&8&9\\
\hline
\rule{0pt}{2ex}   
0&0&3&4&9&6&7&5&8&2&1\\
1&5&1&0&2&8&3&9&6&7&4\\
2&7&6&2&4&1&0&8&9&3&5\\
3&1&5&8&3&7&6&4&0&9&2\\
4&2&9&7&5&4&8&1&3&0&6\\
5&6&7&9&0&3&5&2&4&1&8\\
6&3&8&1&7&5&9&6&2&4&0\\
7&9&4&5&8&2&1&0&7&6&3\\
8&4&0&6&1&9&2&3&5&8&7\\
9&8&2&3&6&0&4&7&1&5&9\\
\end{tabular}
} % end framebox
\myTwoColCaptionvspace[-2.0mm]
\end{table}
}
\fi %return to latex all

\myTwoColSectionvspace[2mm]
On page 38 \mbox{Verhoeff~\wvcite{Verhoeff1969}} continued:
\begin{quote}It is remarkable that up to now no pure decimal codes, with a redundancy of one check digit are known,
which detect all single errors, all transpositions and all twin errors.
\end{quote}
Even a 4-digit code with these properties would imply the existence of 10 pairwise-disjoint 3-digit codes.
The 3 codes presented in Section~\wvref{permfree} fall far short, albeit with some enhanced error detection.
\myTwoColCaptionvspace[-4.0mm]
\begin{table}[!b]
\caption{Detection Conditions for Tables "S" of 3-digit Error Patterns\mydepthbox} \label{ErrorTypes}
\centering
{
\setlength{\tabcolsep}{\mytabcolsepA}
\begin{tabular}{@{}l|*{4}{c|}}
\hline
\rule{0pt}{2.2ex}  
&Error Types
&All Rows
&All Columns
&Main Diagonal \\[\myrowskip]
\iflatexml %only
\hline
\hline
\else %not latexml
\hhline{~|=|=|=|=|}
\fi %return to latex all
\rule{0pt}{2.2ex}   
&Single(*)&$\neg \; abc\leftrightarrow abd$&$\neg \; acd\leftrightarrow bcd$& \\
&&permutation&permutation&  \\[1mm]
\hline 
\rule{0pt}{2.2ex} 
&Transposition&$\neg \; abc\leftrightarrow acb$&$\neg \: abc\leftrightarrow bac$& \\
&&No 2-cycles&No 2-cycles&   \\[1mm]
\hline
\rule{0pt}{2.2ex} 
&Twin&$\neg \; abb\leftrightarrow acc$&$\neg \;aac\leftrightarrow bbc$& \\
& &$\leq 1$ fixed point&$\leq 1$ fixed point&  \\[1mm]
\hline
\rule{0pt}{2.2ex} 
&J. Transposition&  &  & $\neg \; abc\leftrightarrow cab$\\
 &  &  &  &asymmetric \\[1mm]
\hline
\rule{0pt}{2.2ex} 
&J. Twin& &  & $\neg \; aca\leftrightarrow bcb$ \\
&&  &  & permutation \\[1mm]
\hline
\rule{0pt}{2.2ex}  
&Triple&  &  &$\neg \; aaa\leftrightarrow bbb$ \\
&&  & &$ \leq 1$ fixed point \\[1mm]
\hline
\rule{0pt}{2.2ex}  
&Phonetic Left&\multicolumn{3}{c|}{$\forall e: 0\neq S(S(1,c),c) 
\implies \forall x \: \neg\;1xe\leftrightarrow x0c$}\\[1mm]
\hline
\rule{0pt}{2.2ex} 
&Phonetic Right&\multicolumn{3}{c|}{$\forall r: 1\neq S(r,S(r,0)) 
\implies \forall x \: \neg \;r1x\leftrightarrow rx0$}\\[1mm]
\hline
\rule{0pt}{2.2ex} 
&Cyclic&\multicolumn{3}{c|}{ $\forall a, b: b \neq S(S(a,b),a)
\implies \forall x \: \neg \; axb \leftrightarrow xba $}\\[1mm]
\hline
\rule{0pt}{2.2ex} 
&Permutation&\multicolumn{3}{l|}{ Given codewords \(abc, xyz \) }\\
&Free&\multicolumn{3}{c|}{ \(  	multiset(abc)=multiset(xyz) \implies \; abc = xyz \) } \\[1mm]
\hline
\rule{0pt}{2.2ex} 
&\multicolumn{4}{c|}{(*) $Function: S(r,c) 
\implies \neg \; abd\leftrightarrow acd$}\\[1mm]
\hline
\end{tabular}}\end{table}
\myTwoColCaptionvspace[3.0mm]
\section{Disjoint Permutation-Free Decimal Codes}\label{permfree}
\myTwoColCaptionvspace[1.0mm]
The three codes shown in Table~\wvref{DisjointMar2025} 
are pairwise disjoint and each provides protection against all the  error types listed in 
Table~\wvref{ErrorTypes}.
Within each code, the knowledge of the multiset of digits (with multiplicity) in a codeword is sufficient to determine the codeword.
This permutation-free property implies all the properties of Table~\wvref{ErrorTypes}
except for triple error and phonetic error detection.
\iflatexml %only
The codes shown do, however, each protect against all triple and 
phonetic errors\textsuperscript{\wvref{endnote1}}\mydepthbox.
No knowledge beyond this point in the manuscript is required to use the codes.
\else %not latexml
The codes shown do, however, each protect against all triple and phonetic errors\footnotemark.
No knowledge beyond this point in the manuscript is required to use the codes.
\fi %return to latex all

One of the three tables can suffice, if desired, as the codewords in each are a rotation of the codewords in the other two.
For example, note the that code words (100), (010) and (001) are found in Table~\wvref{DisjointMar2025} (a), (b) and (c) respectively.
The subscripts annotating the codewords were used in the construction of the codes
and are explained in Section~\wvref{construction}.
\iflatexml %only
\else %not latexml
\afterpage{\setcounter{footnote}{1}\footnotetext{A permuation-free code with triple errors appears in \wvcite{dunningMar2025}.
It requires only the knowledge of the set of digits, not the multiset, to determine the codeword. 
Alas, the code of Table~\wvref{DisjointMar2025}~(b) contains codewords (5 8 8) and (8 5 5).}}
\fi %return to latex all
\iflatexml %only
\def\cw{5mm}
\begin{table}[!h]
\myTwoColTableLiftvspace
\caption{The Disjoint Decimal Codes \mydepthbox}\label{DisjointMar2025}
\vspace{-2mm}
\myTwoColCaptionvspace
\setlength{\tabcolsep}{0.5em}
\begin{subtable}{.8\textwidth}
\caption{Rotated Left  \mydepthbox}
\begin{tabular}{c|p{\cw}p{\cw}p{\cw}p{\cw}p{\cw}p{\cw}p{\cw}p{\cw}p{\cw}p{\cw}}
$S$&0&1&2&3&4&5&6&7&8&9\\
\hline
\rule{0pt}{2ex}   
0&2& 7& 4& 8& 0& 6& 3& 5& 1& 9\\
1&0& 8& 5& 1& 9& 3& 4& 6& 2& 7\\
2&9& 1& 7& 4& 5& 0& 2& 8& 3& 6\\
3&1& 9& 2& 5& 6& 4& 8& 3& 7& 0\\
4&6& 5& 8& 3& 1& 7& 9& 2& 0& 4\\
5&3& 0& 6& 2& 4& 9& 7& 1& 5& 8\\
6&8& 6& 3& 7& 2& 5& 0& 9& 4& 1\\
7&7& 3& 1& 0& 8& 2& 6& 4& 9& 5\\
8&5& 4& 0& 9& 3& 8& 1& 7& 6& 2\\
9&4& 2& 9& 6& 7& 1& 5& 0& 8& 3\\
\end{tabular}
\end{subtable}
\begin{subtable}{.8\textwidth}
\caption{From Search \mydepthbox}
\begin{tabular}{c|p{\cw}p{\cw}p{\cw}p{\cw}p{\cw}p{\cw}p{\cw}p{\cw}p{\cw}p{\cw}}
$S$&0&1&2&3&4&5&6&7&8&9\\
\hline
\rule{0pt}{2ex}   
0&1$_c$& 3& 0$_l$& 5& 9& 8& 4& 7$_r$& 6& 2\\
1&5& 2$_c$& 9& 7& 8& 4& 6$_r$& 0& 1$_l$& 3\\
2&8& 7& 3$_c$& 6& 0& 1& 5& 2$_l$& 4& 9$_r$\\
3&7& 1$_r$& 5& 4$_c$& 2& 3$_l$& 9& 6& 0& 8\\
4&0$_r$& 4$_l$& 6& 8& 5$_c$& 2& 3& 9& 7& 1\\
5&2& 9& 7& 1& 3& 6$_c$& 0& 4& 8$_r$& 5$_l$\\
6&6$_l$& 8& 2$_r$& 0& 1& 9& 7$_c$& 5& 3& 4\\
7&9& 5& 4& 3$_r$& 7$_l$& 0& 1& 8$_c$& 2& 6\\
8&4& 0& 1& 2& 6& 5$_r$& 8$_l$& 3& 9$_c$& 7\\
9&3& 6& 8& 9$_l$& 4$_r$& 7& 2& 1& 5& 0$_c$\\
\end{tabular}
\end{subtable}
\begin{subtable}{.8\textwidth}
\caption{Rotated Right \mydepthbox}
\begin{tabular}{c|p{\cw}p{\cw}p{\cw}p{\cw}p{\cw}p{\cw}p{\cw}p{\cw}p{\cw}p{\cw}}
$S$&0&1&2&3&4&5&6&7&8&9\\
\hline
\rule{0pt}{2ex}   
0&4& 0& 5& 9& 8& 1& 6& 3& 2& 7\\
1&8& 3& 1& 0& 4& 7& 9& 2& 6& 5\\
2&0& 8& 6& 2& 7& 3& 4& 5& 9& 1\\
3&6& 5& 8& 7& 3& 0& 2& 1& 4& 9\\
4&2& 6& 3& 5& 9& 4& 8& 7& 1& 0\\
5&7& 2& 4& 3& 1& 8& 5& 9& 0& 6\\
6&5& 7& 9& 4& 0& 2& 1& 6& 8& 3\\
7&1& 9& 2& 8& 5& 6& 3& 0& 7& 4\\
8&3& 1& 7& 6& 2& 9& 0& 4& 5& 8\\
9&9& 4& 0& 1& 6& 5& 7& 8& 3& 2\\
\end{tabular}
\end{subtable}
\end{table}
\else %not latexml 
\fi %return to latex all
\myTwoColSectionvspace[-6mm]
\section{Construction of the Disjoint Codes}\label{construction}
The codes here were designed to not have any triple codewords.
To find the code shown in Table~\wvref{DisjointMar2025}~(b) the codewords were separated into two classes.
Those with three different entries (720 possibilities) are denoted by $T$ 
and those with two different entries (270 possibilities) by $D$.
There will be 30 codewords of the forms $r$ (x a a), $c$ (a x a) and $l$ ( a a x) where $a\neq x$ chosen,
and consequently 70 of the form (a b c) of cardinality 3, $|\{a, b, c\}|=3$.
An initial search for a partial code or ``\textit{skeleton}'' with 30 codewords to form $D$ 
satisfying all the conditions of Table~\wvref{ErrorTypes} was conducted first giving 
the permutations denoted by subscripts in Table~\wvref{DisjointMar2025} (b).

The partial code $D$ was fixed and augmented by choosing 70 additional codewords to form $T$.
Let the integer set/sequence
$\{a, a{+}1, \ldots, b\}$ will be denoted by $I_a^b$ or simply $I_n$ for $I_0^{n-1}$.
Let $C = \{\, C_i \;|\; i\in\,I_{120} \}$ represent the collection of the 
$120$ \emph{sets} that are combinations produced by taking subsets of the ten symbols $I_{10}$ three at a time. 
The choice of members of the entries $T$ will then be required to satisfy:
\myTwoColSectionvspace[-1mm]
\begin{equation*}
\begin{aligned}
	&|T \cap C_i| \le 1 \,\forall \, i \in I_{120} 
\end{aligned}
\end{equation*}
This will guarantee that any two of these codewords differ as sets.
However, to guarantee that a latin square is produced,  each of the rows and columns must 
contain all the symbols as reflected in the conditions:
\myTwoColSectionvspace[-1mm]
\begin{equation*}
\begin{aligned}
	& \forall \,r, s \in I_{10} \; | \{ (r\,s\,c_i)\; \ni\, i \in I_{10} \land \,(r\,s\,c_i)\,{\in}\,D\cup T\} |  = 1  \\
	& \forall \,s, c \in I_{10} \; | \{ (r_i\,s\,c)\; \ni\, i \in I_{10} \land \,(r_i\,s\,c)\,{\in}\,D\cup T\} |  = 1  
\end{aligned}
\end{equation*}
% \myTwoColSectionvspace[-1mm]
These conditions were translatable into a mixed integer linear program and were
presented to Sage~\wvcite{sage09} resulting in the desired permutation-free code 
within a few minutes.
Constraints to detect phonetic errors in the code and its rotations were added:
\begin{equation*}
\begin{aligned}
	& \forall \,r, c_i \in I_2^9 \; |\{ (1\,r\,c_i), (r\,0\,c_i) \}|\; \le 1 \; (\textit{left~phonetic}) \\
	& \forall \,r_i, c \in I_2^9 \; |\{ (r_i\,1\,c), (r_i\,c\,0) \}|\; \le 1 \; (\textit{right~~phonetic})\\
	& \forall \,r_i, s \in I_2^9 \; |\{ (r_i\,s\,1), (0\,s\,r_i) \}|\; \le 1 \; (\textit{end~~phonetic})
\end{aligned}
\end{equation*}
The resulting codes of Table~\wvref{DisjointMar2025}~(a-c) are free of phonetic errors and
each individually handles all error types mentioned in this paper.
Should there be data on errors to test these codes against, the properties of the three codes
espoused here will be unaffected by applying any permutation of the digits $I_2^9$ to the triplets. 
The accompanying ``.tex" file ends with ``.html" source to check these codes without regenerating them.
\vspace{0mm}

%\enlargethispage{\equalizelastpgcols}

\iflatexml %only
\else %not latexml 
\def\cw{3.0mm}
\begin{table}[h]
\myTwoColTableLiftvspace
\caption{The Disjoint Decimal Codes\mydepthbox}\label{DisjointMar2025}
%\vspace{2mm}
\myTwoColCaptionvspace[-2mm]
\setlength{\tabcolsep}{0.5em}
\begin{center}
\framebox(210,102){\par
%\begin{tabular}{c|cccccccccc}
\begin{tabular}{c|p{\cw}p{\cw}p{\cw}p{\cw}p{\cw}p{\cw}p{\cw}p{\cw}p{\cw}p{\cw}}
$S$&0&1&2&3&4&5&6&7&8&9\\
\hline
\rule{0pt}{2ex}   
0&2& 7& 4& 8& 0& 6& 3& 5& 1& 9\\
1&0& 8& 5& 1& 9& 3& 4& 6& 2& 7\\
2&9& 1& 7& 4& 5& 0& 2& 8& 3& 6\\
3&1& 9& 2& 5& 6& 4& 8& 3& 7& 0\\
4&6& 5& 8& 3& 1& 7& 9& 2& 0& 4\\
5&3& 0& 6& 2& 4& 9& 7& 1& 5& 8\\
6&8& 6& 3& 7& 2& 5& 0& 9& 4& 1\\
7&7& 3& 1& 0& 8& 2& 6& 4& 9& 5\\
8&5& 4& 0& 9& 3& 8& 1& 7& 6& 2\\
9&4& 2& 9& 6& 7& 1& 5& 0& 8& 3\\
\end{tabular}
\rotatebox[origin = c]{-90}{\sc (a) Rotated Left}\hspace{2mm}
% } % end par
} % end framebox
\end{center}

\begin{center}
\framebox(210,102){
% \begin{tabular}{c|cccccccccc}
\begin{tabular}{c|p{\cw}p{\cw}p{\cw}p{\cw}p{\cw}p{\cw}p{\cw}p{\cw}p{\cw}p{\cw}}
$S$&0&1&2&3&4&5&6&7&8&9\\
\hline
\rule{0pt}{2ex}   
0&1$_c$& 3& 0$_l$& 5& 9& 8& 4& 7$_r$& 6& 2\\
1&5& 2$_c$& 9& 7& 8& 4& 6$_r$& 0& 1$_l$& 3\\
2&8& 7& 3$_c$& 6& 0& 1& 5& 2$_l$& 4& 9$_r$\\
3&7& 1$_r$& 5& 4$_c$& 2& 3$_l$& 9& 6& 0& 8\\
4&0$_r$& 4$_l$& 6& 8& 5$_c$& 2& 3& 9& 7& 1\\
5&2& 9& 7& 1& 3& 6$_c$& 0& 4& 8$_r$& 5$_l$\\
6&6$_l$& 8& 2$_r$& 0& 1& 9& 7$_c$& 5& 3& 4\\
7&9& 5& 4& 3$_r$& 7$_l$& 0& 1& 8$_c$& 2& 6\\
8&4& 0& 1& 2& 6& 5$_r$& 8$_l$& 3& 9$_c$& 7\\
9&3& 6& 8& 9$_l$& 4$_r$& 7& 2& 1& 5& 0$_c$\\
\end{tabular}
\rotatebox[origin = c]{-90}{\sc (b) From Search}\hspace{2mm}
} % end framebox
\end{center}

\begin{center}
\framebox(210,102){
% \begin{tabular}{c |cccccccccc}
\begin{tabular}{c|p{\cw}p{\cw}p{\cw}p{\cw}p{\cw}p{\cw}p{\cw}p{\cw}p{\cw}p{\cw}}
$S$&0&1&2&3&4&5&6&7&8&9\\
\hline
\rule{0pt}{2ex}   
0&4& 0& 5& 9& 8& 1& 6& 3& 2& 7\\
1&8& 3& 1& 0& 4& 7& 9& 2& 6& 5\\
2&0& 8& 6& 2& 7& 3& 4& 5& 9& 1\\
3&6& 5& 8& 7& 3& 0& 2& 1& 4& 9\\
4&2& 6& 3& 5& 9& 4& 8& 7& 1& 0\\
5&7& 2& 4& 3& 1& 8& 5& 9& 0& 6\\
6&5& 7& 9& 4& 0& 2& 1& 6& 8& 3\\
7&1& 9& 2& 8& 5& 6& 3& 0& 7& 4\\
8&3& 1& 7& 6& 2& 9& 0& 4& 5& 8\\
9&9& 4& 0& 1& 6& 5& 7& 8& 3& 2\\
\end{tabular}
\rotatebox[origin = c]{-90}{\sc (c) Rotated Right}\hspace{2mm}
} % end framebox
\end{center}
\myTwoColSectionvspace[-4.0mm]
\end{table}
\myTwoColSectionvspace[-4.0mm]
\fi %return to latex all
\myTwoColSectionvspace[3mm]
\section{Remarks}
\myTwoColSectionvspace[-1mm]
This manuscript shows that mathematicians create results decades after their usefulness has waned.
Dunning \wvcite{dunningMar2024} gives similar permutation-free codes
for other bases as well as a set
\iflatexml %only
\else %not latexml
\newpage\noindent
\fi %return to latex all
of eight comparable length 3 decimal check digit codes
with only the single codeword \texttt{(999)} in common,
which may be potentially useful despite some cyclic and phonetic errors.

Consultation of the lists of references
given by Abdel-Ghaffer \wvcite{Abdel1998} and by Dunning \wvcite{dunningMar2024} is recommended 
when check digit codes with 4 digits or more are needed.
A remarkable sequence of codes for larger bases is given by Damm \wvcite{Damm2007}.
The text by Kirtland \wvcite{kirtland2001} provides a survey and references.

\iflatexml %only
\section{Endnotes}

\begin{enumerate}

\item\label{endnote1}\par{A permuation-free code with triple errors appears in \wvcite{dunningMar2025}.
It requires only the knowledge of the set of digits, not the multiset, to determine the codeword. 
Alas, the code of Table~\wvref{DisjointMar2025}~(b) contains codewords (5 8 8) and (8 5 5).}

\item\label{endnote2}\par{\copyright The author grants copyright permissions as specified in
Creative Commons License CC BY-SA 4.0.  
Attributions should reference the PDF version of this manuscript dated \today, 
which is preferred for printing and redistribution.
% When viewing this HTML version offline, readability may sometimes be improved 
% by temporarily increasing the minimum font size slightly in the web browser's settings.
The author may be reached via e-mail:dunning@bgsu.edu at Larry A. Dunning, Professor Emeritus, 
Department of Computer Science, Bowling Green State University, Ohio 43403, USA.}

\end{enumerate}

\else %not latexml
\fi %return to latex all

\myTwoColSectionvspace[-2mm]
% Generated by IEEEtran.bst, version: 1.13 and \mybstStyle and edited here for space savings

\vfill
\iflatexml %only
% Perl program to copy and run to strip out latexml follows begin{comment}
\begin{comment}
#!/usr/bin/perl -w
# usage > cat ThisFileOnly.tex | striplatexml.pl > Pruned.tex
sub SkipLines {
   while ( $line = <STDIN> ) {
       if ( $line =~ /^\\{1}else %not latexml/ ) { return }; next
   }
}
while ($line = <STDIN>){
    if ( $line =~ /^\\{1}fi %return to latex all/ ) { next };
    if ( $line =~ /^\\{1}iflatexml %only/)          { SkipLines(); next };
    if ( $line !~ /latexml/i ) { print $line };
    next
}
\end{comment}
% Perl code ends with line above end{cooment}
\else %not latexml
\fi %return to latex all
\iflatexml %only
\else %not latexml

\fi %return to latex all

\end{document}